\documentstyle[aps,prb,twocolumn,floats]{revtex}
\input epsf %
\epsfverbosetrue
\newcommand{\tvi}{TVI}
\newcommand{\dof}{DOF}
\newcommand{\w}{\omega}
\begin{document}
\draft
\title{Taber Vibration Isolator for Vacuum and Cryogenic Applications}
\author{H. W. Chan, J. C. Long, and J. C. Price}
\address{Department of Physics - CML, University of Colorado, Boulder,
CO, 80309}
\date{\today}
\maketitle
\begin{abstract}
\begin{center}
{\Large Abstract}
\end{center}
We present a procedure for the design and construction
of a passive, multipole, mechanical high--stop vibration
isolator.  The isolator, consisting of a stack of metal disks connected by
thin wires, attenuates frequencies in the kilohertz
range, and is suited to both vacuum and cryogenic environments.  
We derive an approximate analytical model and compare
its predictions for the frequencies of the normal modes to those of a
finite element analysis.  The analytical model is exact for the modes
involving only motion along and rotation about 
the longitudinal axis, and it gives a good approximate description of the
transverse modes.  These results show that the high--frequency
behavior of a multi--stage isolator is well characterized by the
natural frequencies of a single stage. From the
single--stage frequency formulae, we derive
relationships among the various geometrical parameters of the isolator
to guarantee equal attenuation in all degrees of freedom.  We then 
derive expressions for the attenuation attainable with
a given isolator length, and find that the most
important limiting factor is the elastic limit of the spring wire
material. For our application, which requires attenuations of 250 dB 
at 1 kHz, our model specifies a six--stage design using brass disks
of approximately 2 cm in both radius and thickness, connected by 3 cm
steel wires of diameters ranging from 25 to 75 $\mu$m.
We describe the construction of this
isolator in detail, and compare measurements of the natural
frequencies of a single stage with calculations from the analytical
model and the finite element package.  For translations along and 
rotations about the longitudinal axes, all three results are in
agreement to within 10\% accuracy.
\end{abstract}

\section{Introduction}
A Taber vibration isolator (\tvi) is a passive, multipole, mechanical
high-stop filter for vibration isolation at audio 
frequencies.\cite{Fairbank}  It provides isolation in six degrees 
of freedom and
may reach attenuations of 200--300 dB for all motions.  \tvi s can be
made completely metallic and hence suitable to both cryogenic and 
vacuum environments.

The \tvi\ was invented by R. C. Taber for use with resonant--mass
gravitational wave antennas.\cite{Michelson}  These experiments
typically involve a massive, well--isolated, high--Q resonator with a
fundamental frequency near 1 kHz.  The original application used the
\tvi\ to attenuate vibrations transmitted by wiring leading
to the massive resonator.  In our application, \tvi s are used in an 
apparatus designed to detect gravitational--strength forces between
test masses separated by distances less than 1 mm.\cite{Long,Price}
Specifically, the \tvi s support the test masses, which are 
centimeter--scale, 1 kHz mechanical
oscillators, similar in design to those developed at Bell Labs and
Cornell for use in condensed matter physics experiments.\cite{Kleiman}

Other types of passive, multipole, high-stop vibration isolators have
been described by Aldcroft, {\it et al.}\cite{Aldcroft} and 
Blair {\it et al.},\cite{Blair} also in connection with 
resonant--mass gravitational
wave antennas.  A design involving elastomers for use with laser
interferometric gravitational wave detectors has been described by
Giaime, {\it et al.}\cite{Giaime}

\begin{figure}[hbt]
\begin{center}
\epsfxsize=8.0cm \epsfbox{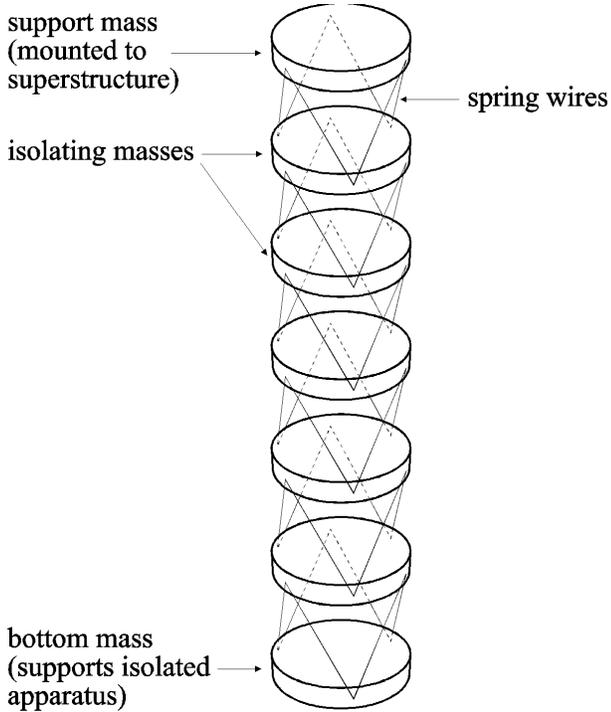}
\caption{Basic geometry of six-stage Taber vibration isolator.}
\label{fig:tvi}
\end{center}
\end{figure}

The basic geometry of the \tvi\ is shown in Fig.~\ref{fig:tvi}.  It
consists of a vertical stack of cylindrical masses connected by
springs made from straight wires under tension.  The hexagonal
arrangement (viewed from above) of the wire attachment points gives
the structure bending stiffness, which raises the frequency of the
pendulum--type modes.  With careful design, this arrangement can yield
approximately equal attenuation in all six degrees of freedom.  For
the particular design developed below, the cylindrical brass masses
are on the centimeter scale, and the springs are made of steel wires
with diameters of tens of microns.

Sec.~\ref{sec:analytical} presents an approximate analytical
model of a \tvi. The model yields simple formulae for the natural
frequencies of an isolator with a single spring--mass pair (or
``stage''), and an accurate solution for the normal modes of a 
complete isolation stack.  Sec.~\ref{sec:ansys} presents a 
numerical analysis of the normal modes of a multi--stage
isolator and shows that the results agree very well with the
analytical model. Sec.~\ref{sec:opt} uses the single--stage natural
frequency formulae to optimize the design of a multi--stage isolator
for uniform attenuation in all degrees of freedom, given the specific
geometrical constraint of finite stack length.  The subsequent
predictions for the number of stages and for the spring wire diameter are
used to construct an actual \tvi, which is described in 
Sec.~\ref{sec:design}. The
measurements of the natural frequencies of a single stage are 
compared with calculations from the analytical model and the 
numerical analysis.

\section{Approximate Analytical Model}
\label{sec:analytical}
\begin{figure}[hbt]
\begin{center}
\epsfxsize=9cm \epsfbox{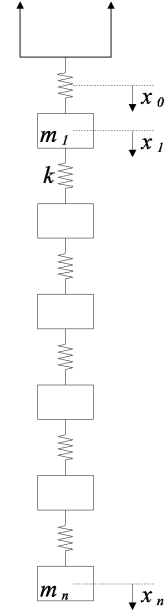}
\caption{One--dimensional lumped--element spring--mass chain; basis of
  simple analytical model of Taber isolator.}
\label{fig:1dA}
\end{center}
\end{figure}

The analytical model of the \tvi\ is based on a combination
of the lumped--element one--dimensional spring--mass chain shown in 
Fig.~\ref{fig:1dA}, and the single stage in Fig.~\ref{fig:1dB}.  This 
model is exact for translations along and rotations about the
vertical axis, and, as shown below, it gives a good approximation 
of the transverse motions as well.

A single stage with one degree of freedom (\dof) is useful for
understanding the attenuation of a multi-stage isolator.  If the 
spring is displaced at the top by a vibration of frequency $\w$ (the 
frequency to be filtered by the isolator, or operating frequency) 
and amplitude $x(\w)$, the
equation of motion for this system may be written:
\begin{equation}
\label{eq:sho}
-m\w^{2}x_{0}(\w) = k[x(\w) - x_{0}(\w)],
\end{equation}
where $m$ is the mass, $k$ is the spring constant, and 
$x_{0}(\w)$ is the amplitude of the suspended mass.
Re-arranging this equation yields the single--stage transfer 
function for one \dof:
\begin{equation}
\label{eq:atten}
T_{0}(\w) \equiv \frac{x_{0}(\w)}{x(\w)} = 
\frac{\w_{0}^{2}}{\w_{0}^{2}-\w^{2}},
\end{equation}
where $\w_{0} = \sqrt{k/m}$ is the natural frequency of the
single--stage isolator.

The rest of this analysis considers operation frequencies well
above the resonant frequencies of the system ($\w \gg \w_{0}$).  In
this regime, $T_{0}(\w) = (\w_{0}/\w)^{2}$. In the case of an isolator
of $n$ identical stages, each with natural frequency $\w_{0}$ 
(Fig.~\ref{fig:1dA}), $x_{\mu}(\w) \gg x_{\mu+1}(\w)$ for the
$\mu$th stage, so that Eq.~\ref{eq:sho} may be applied to each successive 
stage. The displacement of each stage with respect to that immediately above 
is then simply another factor of $(\w_{0}/\w)^{2}$,
so that the attenuation in displacement from the support to the 
$n$th suspended mass, for one \dof, is given by $(\w_{0}/\w)^{2n}$. The 
extremely rapid dependence of the attenuation on frequency is, of
course, the reason why multi-stage isolators are so effective.

The complete expression for the transfer function for an undamped
system of $n$ stages with masses $m_{\mu}$, spring constants $k_{\mu}$, 
and one \dof \ is given by:\cite{Aldcroft}
\begin{equation}
\label{eq:tf}
T_{n}(\w) \equiv \frac{x_{n}(\w)}{x(\w)} = 
\prod_{\mu=\nu=1}^{n}{\frac{k_{\mu}/m_{\mu}}{\Omega_{\nu}^{2}-\w^{2}}}.
\end{equation}
Here, $\sqrt{k_{\mu}/m_{\mu}} = \w_{\mu}$ is the natural frequency 
of the $\mu$th stage, and $\Omega_{\nu}$ is the frequency of 
the $\nu$th normal mode.  In the high 
frequency regime,  this result reduces
to $(\w_{0}/\w)^{2n}$, for an isolator with identical stages 
($\w_{\mu} = \w_{0}$).

In order to use Eq.~\ref{eq:tf} to estimate the total attenuation of
the \tvi \ for each \dof, the analytical model can be used to approximate
the single--stage frequencies $\w_{\mu}$ and to calculate the normal mode 
frequencies $\Omega_{\nu}$.  The first step is to derive the 
linear and torsional spring constants for each \dof.  
\begin{figure}[hbt]
\begin{center}
\epsfxsize=8.5cm \epsfbox{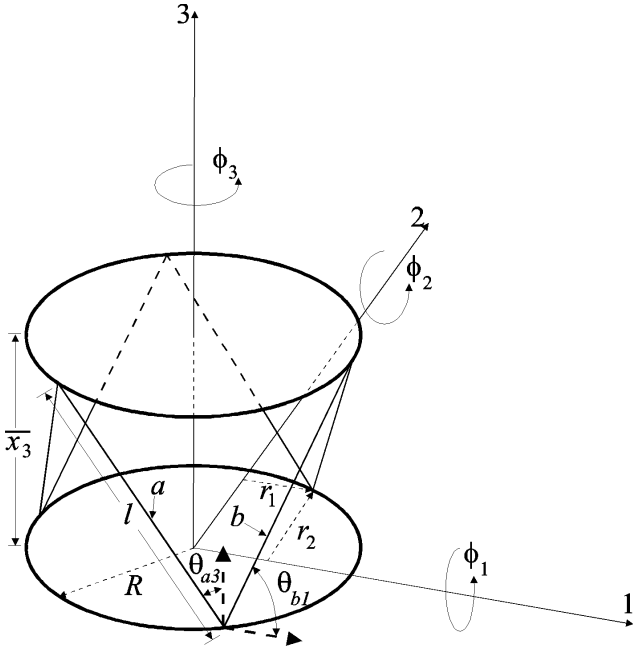}
\caption{Single \tvi \ stage with parameters used for calculation 
  of spring constants.}
\label{fig:1dB}
\end{center}
\end{figure}

\subsection{Spring Constants}
Using the axes defined in Fig.~\ref{fig:1dB}, the normal modes of
the \tvi \ are modeled as pure translations along and rotations 
about the 1-, 2-, and 3-axes.  This is an approximation because in 
reality the 1- and 2- translations and rotations are coupled.
The evaluation of the spring constant for translation along the
3-axis is the most straightforward.  In Fig.~\ref{fig:1dB}, $\theta_{a3}$
is the angle between the 3-axis and wire $a$, $\overline{x_{3}}$ is
the equilibrium spacing between any two stages, $l$ is the length of the
wire, $A$ is the cross-sectional area of the wire, and $E$ is the
modulus of elasticity of the wire.  In this analysis, each of these
parameters is assumed to have a unique value for a particular stage.
Since the basic geometry of all stages is the same, however, the
functional form of the spring constants for each stage is also the same.  
The stage subscript, $\mu$, can therefore be dropped to economize the
notation.  The relationship between an infinitesimal force $dF$ along
a spring wire and displacement $dl$ is then: 
\begin{equation}
dF = -E \frac{A}{l} dl.
\end{equation}
For an initial longitudinal displacement $dx_{3}$, the
displacement along the wire is: 
\begin{equation}
dl = dx_{3} \cos{\theta_{a3}}
\end{equation}
so that: 
\begin{equation}
dF = -E\frac{A}{l}\cos{\theta_{a3}}dx_{3}.
\end{equation}
The longitudinal component of the force is then:
\begin{equation}
dF_{3} = -E \frac{A}{l}\cos^{2}{\theta_{a3}}dx_{3}.
\end{equation}
This expression is equivalent for all six wires per stage, so that the 
total spring constant for motion along the 3-axis is:
\begin{equation}
k_{x_{3}} = 6E \frac{A}{l}\cos^{2}{\theta_{a3}} = 
\frac{6EA(l^{2}-R^{2})}{l^{3}}, 
\end{equation}
where $R$ is the radius of the stage, and we have used $R^{2} = l^{2} -
x_{3}^{2}$ from the hexagonal geometry of the attachment points. In 
general, if $\theta$ is the angle between a wire and the direction
of translation, the corresponding contribution to the spring constant 
for that \dof \ is $k = E A \cos^{2}{\theta} / l$.

Referring again to Fig.~\ref{fig:1dB}, two wires per stage have angle
$(\pi/2) - \theta_{a3}$ with respect to the 1-axis. Their
contribution to the corresponding spring constant is therefore 
$k = 2EAR^{2}/l^{3}$. The remaining four wires each have angle $\theta_{b1}$
with respect to this axis, bringing the total spring constant to 
$k_{x_{1}} = 3EAR^{2}/l^{3}$.

Finally, from the figure, four wires per stage have angle
$\theta_{b2}$ with respect to the $2$-axis, so that their total 
contribution to the spring constant is $k = 3EAR^{2}/l^{3}$.  
This is the complete expression for $k_{x_{2}}$, since the remaining 
two wires are orthogonal to the $2$-axis.  These results are
summarized in Table~\ref{tab:k}.

Computation of the torsional spring constants is simplified by
expressing the infinitesimal torques $d \tau_{j}$ about each axis in 
terms of the infinitesimal displacements $dx_{j}$ of the attachment
points. If $d \phi_{j}$ is the infinitesimal rotation of the effective
lever arm $r_{k}$ about the $j$th axis, we have, for each wire on
the stage:
\begin{equation}
d \tau_{1} = r_{2} dF_{3} = -k_{\tau_{1}} d \phi_{1},
\end{equation}
\begin{equation}
d \tau_{2} = r_{1} dF_{3} = -k_{\tau_{2}} d \phi_{2}.
\end{equation}
Substituting $dF_{3} = -(k_{x_{3}}/6) dx_{3}$, $d \phi_{1} = dx_{3}/r_{2}$, 
and $d \phi_{2} = dx_{3}/r_{1}$ yields:
\begin{equation}
k_{\tau_{1}} = r_{2}^{2} \frac{k_{x_{3}}}{6},\ k_{\tau_{2}} 
= r_{1}^{2} \frac{k_{x_{3}}}{6}.
\end{equation}

From the geometry in Fig.~\ref{fig:1dB}, only four wires contribute to
the rotation of the stage about the 1-axis, each attached at a distance
$r_{2} = R\sqrt{3}/2$ from that axis. The total torsional constant is
therefore:
\begin{equation}
k_{\tau_{1}} = 4 \left(\frac{R \sqrt{3}}{2}\right)^{2} 
\frac{k_{x_{3}}}{6} = 
3EA\frac{R^{2}(l^{2}-R^{2})}{l^{3}}.
\end{equation}
All six wires contribute to the rotation about the 2-axis, with two
attached at a distance $r_{1} = R$ from the axis and the remaining
four at $r_{1} = R/2$:
\begin{equation}
k_{\tau_{2}} = 2 R^{2} \frac{k_{x_{3}}}{6} + 4 \left(\frac{R}{2}\right)^{2}
\frac{k_{x_{3}}}{6} = 3EA\frac{R^{2}(l^{2}-R^{2})}{l^{3}}.
\end{equation}

For the remaining torque about the 3-axis, we consider the force due to
one wire at the attachment point lying on the 1-axis. Here, the force 
is along the 2-axis, so that
\begin{equation}
d \tau_{3} = r_{3} dF_{2} = -k_{\tau_{3}} d \phi_{3}.
\end{equation}
Substituting $dF_{2} = -(k_{x_{2}}/4) dx_{2}$, $d \phi_{3} = dx_{2}/r_{3}$
yields:
\begin{equation}
k_{\tau_{3}} = r_{3}^{2} \frac{k_{x_{2}}}{4}.
\end{equation}
From the symmetry of the stage about the 3-axis, the contribution of
all six wires, each at a distance $r_{3} = R$ from the axis, must be
the same. The total torsional constant is therefore: 
\begin{equation}
k_{\tau_{3}} = 6 R^{2} \frac{k_{x_{2}}}{4} = 
\frac{9}{2}EA\frac{R^{4}}{l^{3}}
\end{equation}
These results are included in Table~\ref{tab:k}.

\begin{table*}
\caption{Single-stage spring constants and natural frequencies for each 
\dof \ in the analytical model.}
\label{tab:k}
\begin{tabular}{lcc}\hline
\multicolumn{1}{c}{\it Motion} & \multicolumn{1}{c}{\it Spring
Constant} & \multicolumn{1}{c}{\it Natural Frequency}\\ \hline
Translation, 1- and 2-axes 
& $k_{x_{1}} = k_{x_{2}} = 3EAR^{2}/l^{3}$ 
& $w_{x_{1}} = w_{x_{2}} = R \sqrt{\frac{3EA}{ml^{3}}}$ 
\\
Translation, 3-axis 
& $k_{x_{3}} = 6EA(l^{2}-R^{2})/l^{3}$ 
& $w_{x_{3}} = \sqrt{\frac{6EA(l^{2}-R^{2})}{ml^{3}}}$ 
\\
Rotation, 1- and 2-axes 
& $k_{\tau_{1}} = k_{\tau_{2}} = 3EAR^{2}(l^{2}-R^{2})/l^{3}$ 
& $w_{\tau_{1}} = w_{\tau_{2}} = 2 \sqrt{\frac{3EA(l^{2}-R^{2})}
{ml^{3}(1+\frac{1}{3}(\frac{t}{R})^{2})}}$ 
\\
Rotation, 3-axis
& $k_{\tau_{3}} = 9EAR^{4}/(2 l^{3})$ 
& $w_{\tau_{3}} = 3 R\sqrt{\frac{EA}{ml^{3}}}$ 
\\
\end{tabular}
\end{table*}

\subsection{Resonant Frequencies}
The natural frequencies for the translational modes of a particular
stage are simply $\w_{x_{j}} = \sqrt{k_{x_{j}}/m}$, 
where $j = 1-3$ refers to each axis and $m$ is the mass 
of the disk.  For the rotational modes, the natural frequencies are 
$\w_{\tau_{j}} = \sqrt{k_{\tau_{j}}/I_{\tau_{j}}}$.
Here, $I_{\tau_{3}} = mR^{2}/2$ is the moment of inertia of the 
disk about the 3-axis, and
$I_{\tau_{1}} = I_{\tau_{2}} = mR^{2}(1+(t/R)^{2}/3)/4$ is the 
moment about the 1- and 2-axes, where $t$ is the disk thickness.
The natural frequencies for each \dof \ are included 
in Table~\ref{tab:k}.

\begin{figure}[hbt]
\begin{center}
\epsfxsize=8.5cm \epsfbox{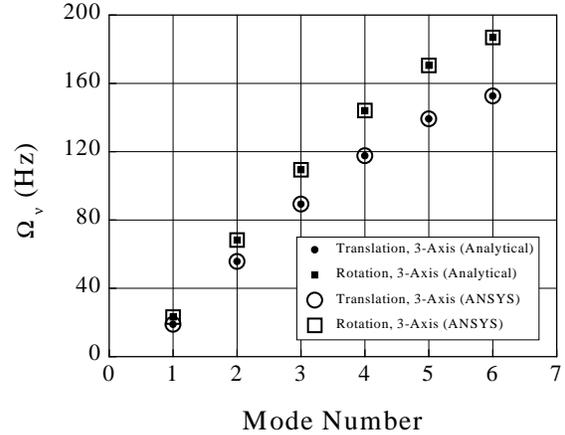}
\vspace{.5cm}
\caption{Resonant frequencies $\Omega_{\nu}$ vs. mode number 
  for translations and rotations (3-axis).}
\label{fig:mod3}
\end{center}
\end{figure}

\begin{figure}[hbt]
\begin{center}
\epsfxsize=8.5cm \epsfbox{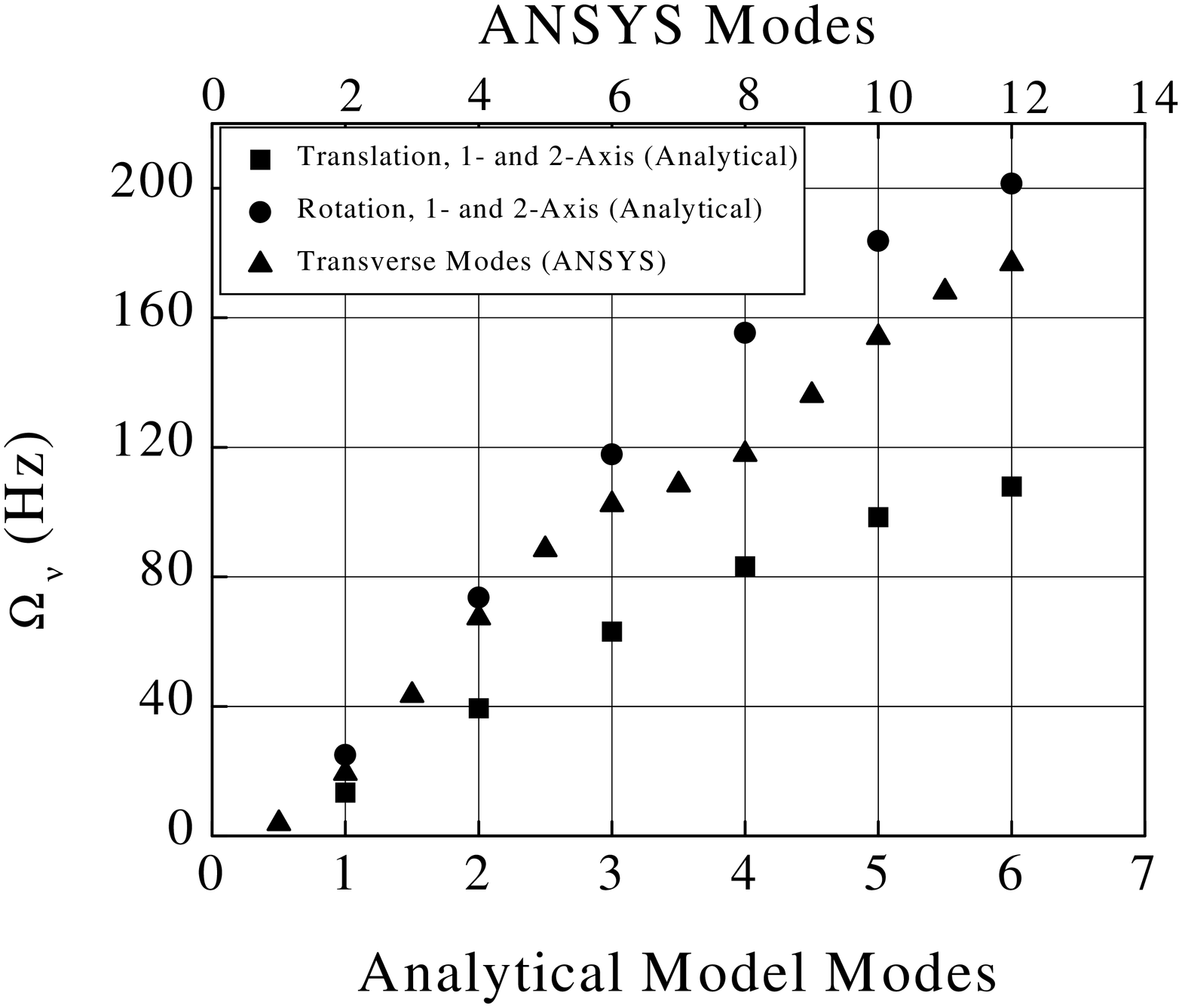}
\vspace{.5cm}
\caption{Resonant frequencies $\Omega_{\nu}$ for translations and
  rotations (1- and 2-axes).  For the numerical
  analysis, frequencies are arranged by mode number (upper axis).  
  For the analytical model, frequencies are arranged in increasing
  order for each \dof.}
\label{fig:mod12}
\end{center}
\end{figure}

For each \dof, the normal mode frequencies $\Omega_{\nu}$ must be
calculated in a complete solution to the one--dimensional
lumped--element spring--mass chain in Fig.~\ref{fig:1dA}. The choice 
of six stages for the model, while technically arbitrary at 
this point, is motivated by the results of the design optimization in 
Sec.~\ref{sec:opt}.  The analytical solution leads to a twelfth--order
polynomial in $\w$ for translation or rotation in the relevant
coordinate.  The normal mode frequencies $\Omega_{\nu}$ are computed by
substituting the expressions for the spring constants in
Table~\ref{tab:k} into the characteristic equation and finding the
roots numerically.  The parameters entering into the characteristic
equation (mass radii and thickness, wire length, diameter, and 
elastic modulus) also derive from the results of the optimization procedure in 
Sec.~\ref{sec:opt}, and are listed in Tables~\ref{tab:construction} 
and~\ref{tab:constructiond}.
The frequencies are plotted in Figs.~\ref{fig:mod3} and ~\ref{fig:mod12}.
The twelve translational frequencies and twelve rotational frequencies
for the 1- and 2-axes are two--fold degenerate.

\begin{figure}[hbt]
\begin{center}
\epsfxsize=8.5cm \epsfbox{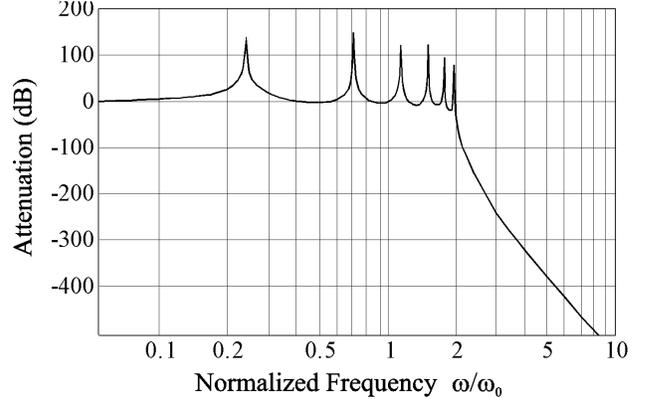}
\caption{Transfer function vs. operation frequency for 
  six--stage isolator.}
\label{fig:tf}
\end{center}
\end{figure}

Fig.~\ref{fig:tf} shows the transfer function evaluated for the case
of six stages and $\w_{\mu} = \w_{0}$ for each stage.  The highest
normal mode frequency is close to $2 \w_{0}$
(this would be exact for an infinite chain), and the asymptotic
attenuation is nearly reached at $\w \approx 3 \w_{0}$.  

The single--stage natural frequency $\w_{0}$ nearly completely
characterizes the behavior of a multi-stage \tvi \ with identical
stages.  This frequency depends on the design parameters of the
\tvi \ in a simple and explicit way, and therefore can be used directly
for design optimization.  Before discussing optimization, the
analytical model is compared to a numerical analysis.
\vspace{1cm}
\vspace{1.1cm}

\section{Finite Element Analysis}
\label{sec:ansys}
A numerical model of a six stage \tvi \ is constructed using
the ANSYS\cite{ansys} finite element analysis software
package.  Rigid bodies with six \dof \ model the brass disks,
and elastic beams, also with six \dof, model the steel wires. The
beams are arranged in the geometry of the wires in Fig.~\ref{fig:1dB},
and are coupled rigidly at their endpoints to the point masses representing
the disks.  In addition to mass, the required input 
parameters for the point masses include the moments of inertia for
each rotational \dof.  The required input parameters for the beams 
include density and modulus of elasticity.  All input parameters are
derived from the data in Tables~\ref{tab:construction} 
and~\ref{tab:constructiond}. 

The frequencies of the 36 normal modes calculated by ANSYS are
compared to the frequency calculations from the analytical model in 
Figs.~\ref{fig:mod3} and ~\ref{fig:mod12}.  The results in 
Fig.~\ref{fig:mod3} show that the analytical model is exact 
for translations along and rotations about the 3-axis.  
In Fig.~\ref{fig:mod12}, remaining frequencies
calculated in ANSYS are arranged by mode number (upper axis).  
The results from the analytical model are arranged by 
increasing frequency for each \dof. The plot 
illustrates that the true normal mode frequencies fall between the
uncoupled translation and rotation frequencies found analytically.

This analysis suggests that the approximate analytical model of the \tvi \ 
is sufficiently accurate for use in the development of a
working design.  Furthermore, for an isolator with identical stages
operating in the high frequency domain, the single--stage resonant 
frequency $\w_{0}$ for each \dof \ from the model is a sufficient
parameter with which to optimize the design.

\section{Design Optimization}
\label{sec:opt}
From Eq.~\ref{eq:tf}, the attenuation for a multi-stage isolator is most
strongly dependent on the number of stages, $n$.  While it is
important to choose a sufficient number of stages for the degree of
attenuation desired, the maximization of $n$ given the geometrical
constraints of the system must be balanced with at least two other
important factors.  First, attention should be given to the extent to
which attenuation is desired in each \dof \ of the isolated system.
Second, it is essential that the transverse vibrational
modes of the spring wires be kept well above the operational
frequency, $\w$, so that the wires function as simple springs. 

\subsection{Uniform Isolation in each \dof}
\label{ssec:uni}
The particular application in our laboratory requires the isolation of
vibrations in the kilohertz range in all \dof.  The normal modes 
for each \dof \ of the \tvi \ should therefore be essentially 
the same.  Using the single--stage analytical model, relations 
between the geometrical parameters of the \tvi \ can be found by 
equating the natural frequencies for each \dof.  

From the expressions in Table~\ref{tab:k}, the natural frequencies 
$\w_{x_{1,2}}$ and $\w_{\tau_{3}}$ always differ by a factor of
$\sqrt{3}$. Taking an approximate average of these two terms and
equating it to the other natural frequencies in the table yields:
\begin{equation}
\frac{6EAR^{2}}{ml^{3}} \approx
\frac{6EA(l^{2}-R^{2})}{ml^{3}} \approx
\frac{12EA(l^{2}-R^{2})}{ml^{3}(1+\frac{1}{3}
(\frac{t}{R})^{2})}.
\end{equation}
Simplifying:
\begin{equation}
R^{2} \approx (l^{2}-R^{2}) \approx 
\frac{2(l^{2}-R^{2})}{(1+\frac{1}{3}(\frac{t}{R})^{2})}.
\end{equation}
Equating the first two terms gives the relation $R = l/\sqrt{2}$. 
The second two terms yield $t = R\sqrt{3}$, and the remaining equality
yields no new information.  In terms of the wire length $l$, the disk
radius and thickness are related by
\begin{equation}
R = l/\sqrt{2},\ t = l\sqrt{3/2}.
\label{eq:lconv}
\end{equation}
Ideally, Eq.~\ref{eq:lconv} guarantees equal single--stage frequencies and
therefore equal attenuation of vibrations for each \dof \ of the
multi--stage \tvi\ with identical stages.  In practice this condition
is relaxed somewhat, as explained in Sec.~\ref{sec:design} and 
reflected in Figs.~\ref{fig:mod3} and ~\ref{fig:mod12}.

\subsection{Optimal Number of Stages}
\label{ssec:num}
The maximum attenuation per \dof \ can now in principle be obtained
by maximizing the number of stages $n$, subject to the above
constraints.  At this point, however, the design is of course limited
by the geometrical constraints of the containment system and the
properties of appropriate \tvi \ construction materials.

With the equality of the natural frequencies of each \dof \ guaranteed by the
constraints in the previous section, the attenuation may optimized
based on motion in only one particular \dof. For the choice of 
longitudinal motion along the 3-axis, the single-stage transfer 
function is given by:
\begin{equation}
T_{0x_{3}} = \frac{\w_{0}^{2}}{\w^{2}} = \frac{1}{\w^{2}} 
\frac{6EA(l^{2}-R^{2})}{ml^{3}}.
\end{equation}
Expressing the disk mass $m$ in terms of the density $\rho_{d}$ and
the dimensions of the disk yields:
\begin{equation}
T_{0x_{3}} = \frac{1}{\w^{2}}\frac{6EA(l^{2}-R^{2})}
{\pi \rho_{d} R^{2} t l^{3}},
\end{equation}
and substituting Eq.~\ref{eq:lconv} gives:
\begin{equation}
T_{0x_{3}} = \frac{1}{\w^{2}} 
\frac{12EA}{\sqrt{3}\pi \rho_{d} l^{4}}.
\label{eq:t31}
\end{equation}

The geometrical constraint of the containment system is modeled
by limiting the total possible height of the stack to
some finite value.  This translates into a total finite length $L$ for
all wires along one side of the stack.  With $n$ total stages of equal
length, the wire length $l$ per stage is $l = L/n$.

The cross-sectional area $A$ of the spring wires is limited by the
elastic limit stress $s$, or the stress at which acoustic emission 
becomes intolerable. 
If $F$ is the force on one of the six wires supporting a
single--stage isolator, the minimum cross-section is given by
\begin{equation}
A_{0} = \frac{F}{s} = \frac{\pi R^{2} t \rho_{d} g}{6s}.
\end{equation}

As this point, the requirement that each stage in the multi-stage
\tvi\ model have equivalent geometry is relaxed in order to
permit the minimum wire cross--section per stage.  This results in a
range of natural frequencies $\w_{\mu}$ in the stack.  However, as
long as Eq.~\ref{eq:lconv} is made to hold for each stage, and as long as
$\w_{\mu} \ll \w$ for all $\mu$, the multi--stage \tvi \ will still 
operate effectively in all \dof.

For the $\mu$th stage in a stack of $n$ stages (with $\mu = 1$
corresponding to the top stage), the minimum wire cross section for
that stage requires a factor of $(n-\mu + 1)$ due to the weight of the
stages below it. Substituting Eq.~\ref{eq:lconv}, 
\begin{equation}
A_{\mu} = \frac{\pi l^{3} \rho_{d} g (n-\mu + 1)}{4 \sqrt{6}s}.
\label{eq:crossec}
\end{equation}
Ensuring the maximum tolerable force on each wire also
has the effect of maximizing the frequencies of the transverse modes
on the spring wires, an important point to be considered below.

\begin{figure}[hbt]
\begin{center}
\epsfxsize=8.5cm \epsfbox{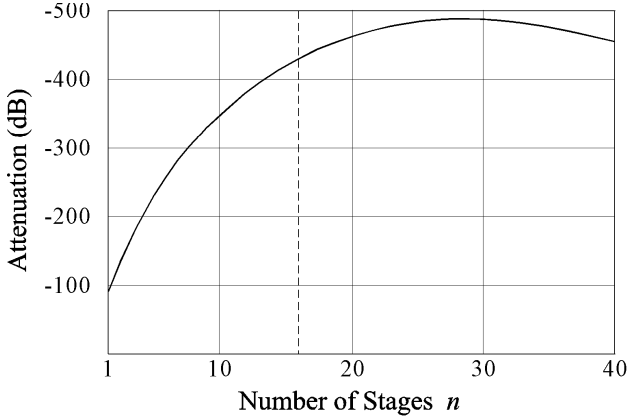}
\caption{Attenuation ($T$) in dB vs. number of stages for analytical model of a
  \tvi\ with operation frequency of 1 kHz and properties in
  Table~\ref{tab:construction}. The
  dashed line indicates the approximate limit of the asymptotic
  regime. Scale: 1 dB = 20 $\log{T}$.} 
\label{fig:attenuation}
\end{center}
\end{figure}

Inserting the last expression into Eq.~\ref{eq:t31}, and again using
Eq.~\ref{eq:lconv}, the transfer function for the $\mu$th stage of
the $n$-stage isolator is given by:
\begin{equation}
T_{\mu x_{3}} = \frac{\w_{\mu}^{2}}{\w^{2}} = \frac{n}{\w^{2}} 
\frac{Eg(n-\mu + 1)}{\sqrt{2}sL}.
\label{eq:transi}
\end{equation}
The complete transfer function for the $n$-stage isolator is then the
product of the factors for each stage:
\begin{eqnarray}
T_{x_{3}} & = & \left(\frac{n}{\w^{2}}\frac{Egn}{\sqrt{2}sL}\right)
\left(\frac{n}{\w^{2}}\frac{Eg(n-1)}{\sqrt{2}sL}\right)
\left(\frac{n}{\w^{2}}\frac{Eg(n-2)}{\sqrt{2}sL}\right) \nonumber \\ 
& & \times \cdot \cdot \cdot \times 
\left(\frac{n}{\w^{2}}\frac{Eg}{\sqrt{2}sL}\right)
= \left(\frac{n}{\w^{2}}\frac{Eg}{\sqrt{2}sL}\right)^{n} n! 
\end{eqnarray}
This expression illustrates importance of using spring materials
of low elastic modulus and high elastic limit.
Fig.~\ref{fig:attenuation} shows a plot of $T_{x_{3}}$, using the
parameters in Table~\ref{tab:construction}.  The constraints specified
in this model limit the exponential rise in attenuation with
the number of stages ($T \approx (\w_{1}/\w)^{2n}$).  
At first glance, one might expect to attain a
maximum attenuation of about 500 dB with 30 stages (1 dB $\equiv
20 \log{T}$).  However, this
analysis applies only to the asymptotic regime, the upper limit of
which, from Fig.~\ref{fig:tf}, is defined by $\w \simeq 3\w_{0}$.
From Eq.~\ref{eq:transi}, the highest $\w_{\mu}$ in the model
corresponds to $\mu = 1$:
\begin{equation}
w_{1} = \sqrt{\frac{n^{2}Eg}{\sqrt{2}sL}}.
\end{equation}
Requiring $\w_{1} < \w/3$ in this expression yields $n < 16$,
suggesting that attenuations above 400 dB may still be possible.
 
\subsection{Effect of Spring Transverse Modes}
\label{ssec:trans}
In the design of a \tvi, care must be taken to ensure that the
frequencies of the transverse modes of the spring wires lie well above
the operation frequency.  Otherwise, the wires will not act as
lumped--element springs.

\begin{figure}[hbt]
\begin{center}
\epsfxsize=8.5cm \epsfbox{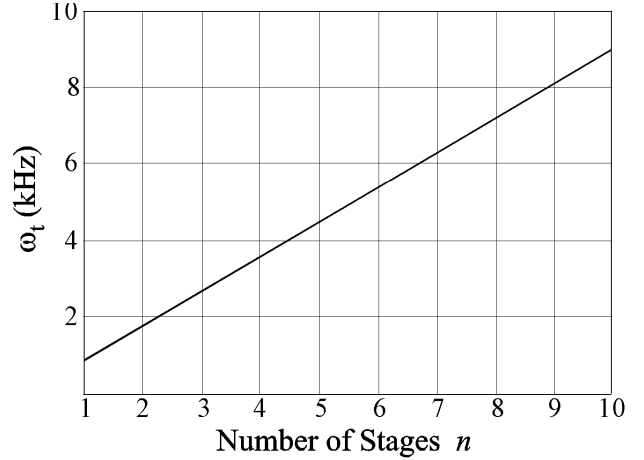}
\caption{Frequency of spring wire fundamental transverse mode $\w_{t}$
  vs. number of stages for analytical model of a \tvi\ with operation 
  frequency of 1 kHz and properties in Table~\ref{tab:construction}.}
\label{fig:transverse}
\end{center}
\end{figure}

The fundamental transverse frequency of a wire of length $l$ is given by:
\begin{equation}
\w_{t} = \frac{\pi}{l}\sqrt{\frac{T}{\eta}},
\end{equation}
where $T$ is the tension in the wire and $\eta$ is the wire mass per
unit length.  From the previous section, the tension in the wire is
automatically maximized; it is just the elastic limit times the cross-section: 
$T = sA$. The mass per unit length is $\eta = \rho_{w} A$, where
$\rho_{w}$ is the wire mass density.  Applying Eq.~\ref{eq:lconv}, the
lowest transverse frequency becomes:
\begin{equation}
\w_{t} = \frac{n \pi}{L}\sqrt{\frac{s}{\rho_{w}}}.
\end{equation}
This function is plotted in Fig.~\ref{fig:transverse}, using the
parameters in Table~\ref{tab:construction}.  For this model, 
any choice of $n$ greater than 3 insures that all transverse 
spring frequencies are at least twice the operational frequency.

An important assumption made in the optimization procedure is that the
diameter of the spring wires is allowed to decrease on successively lower
stages in the stack.  From Eq.~\ref{eq:crossec} for the
cross-sectional area of the spring wires on a given stage, and using
Eq.~\ref{eq:lconv}, the diameter of the spring wires on the $\mu$th
stage of an $n$-stage \tvi \ is given by:
\begin{equation}
d_{w} = \sqrt{\frac{L^{3} \rho_{d}g(n-\mu +1)}{\sqrt{6}n^{3}s}}. 
\label{eq:diameter}
\end{equation}
The diameter is plotted as a function of stage number in 
Fig.~\ref{fig:wire}, for the cases $\mu = 1$
(top stage) and $\mu = n$ (bottom stage), using the parameters in
Tables~\ref{tab:construction} and~\ref{tab:constructiond}.  
The values of the curves at a given $n$
define the range of wire diameters needed for the construction of a \tvi
\ with the attenuation shown in Fig.~\ref{fig:attenuation}.  

For $n > 10$, the diameters required fall well below 25 $\mu$m,
which may be a practical limit.  Note the dependence of 
Eq.~\ref{eq:diameter} on $\rho_{d}$,
illustrating the importance of choosing stage masses of high density.
Wires of larger diameter may be used at some cost of attenuation, but 
care should be taken to avoid transverse modes in the operating range.

\section{Construction and Test of Six--Stage Isolator}
\label{sec:design}
\subsection{Choice of Materials}
The preceding discussion mandates a choice of a high--density
material for the construction of the \tvi \ masses.  For our specific
application, the masses must non-magnetic, and be vacuum and
cryogenically compatible.  An ideal choice would be tungsten, but
brass is selected for its low cost and ease of machining.

The spring wires should have a combination of low elastic modulus and
density, and, most importantly, high elastic limit. Beryllium copper
and certain aluminum alloys have ideal characteristics, and have
the added advantage of being non-magnetic. However, type 304
stainless steel is chosen for its availability at low cost in many
different diameters.\cite{wire}

The construction of a \tvi \ for a small laboratory application is 
considerably simplified if the spring wires can be made to attach to 
a set of coplanar screws bolted to the outer rim of each suspended
disk, as suggested by Fig.~\ref{fig:tvi}. This makes impractical the 
requirement $t > l$, as derived in Sec.~\ref{ssec:uni}.  If the screws
are set into the midplane of each disk, the hexagonal geometry of the
attachment points guarantees the stage spacing to be $R$, as long as
$R = l/\sqrt{2}$.  
\begin{figure}[hbt]
\begin{center}
\epsfxsize=8.5cm \epsfbox{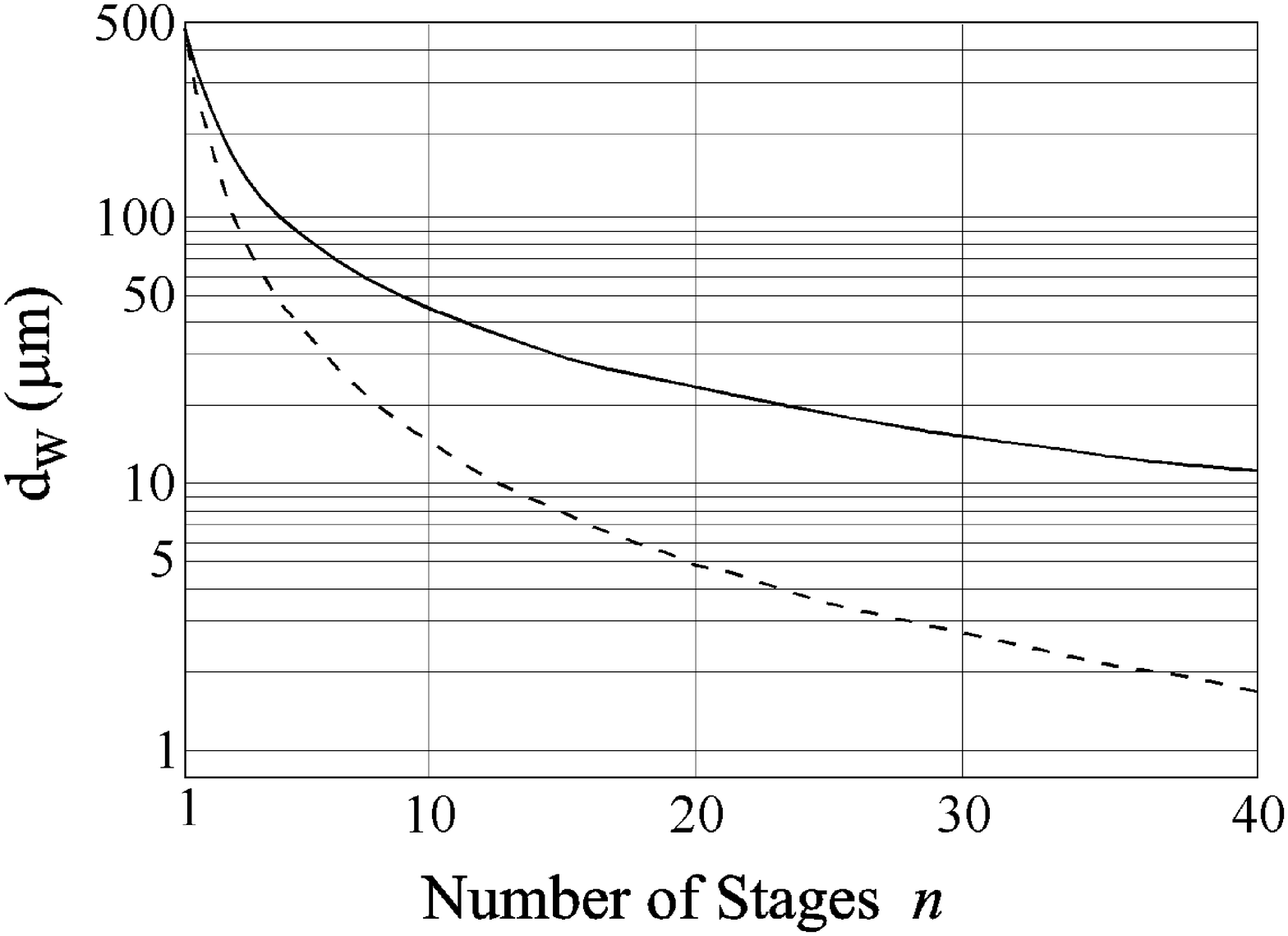}
\caption{Wire diameter vs. number of stages for top stage (solid
  curve) and bottom stage (dashed curve) in analytical model of a
  \tvi\ with operation frequency of 1 kHz and properties in
  Tables~\ref{tab:construction} and ~\ref{tab:constructiond}.}
\label{fig:wire}
\end{center}
\end{figure}

A simple choice would be to set both the stage thickness
and the inter-stage gap to $R/2$.  However, smaller disk
thicknesses drive the frequencies of the 1- and 2-rotational modes
higher relative to the other modes, so that  
$t \approx R/\sqrt{2} \approx l/2$ is a better choice. 
The rotational modes increase, but
not enough to reach into the domain of the 1 kHz operating frequency 
(as seen in Fig.~\ref{fig:mod12}).  In fact, this choice changes 
the optimization expressions used in Secs.~\ref{ssec:num} 
and~\ref{ssec:trans} only by small numerical factors.

The size of our vacuum chamber limits the total height of the stack to
11.5 cm.  With a stage separation distance $R$, the total wire
length is then simply $L = R\sqrt{2}$, or 16.2 cm.

The total wire length needed and the properties of the mass and
spring wire materials are listed in Tables~\ref{tab:construction} and 
~\ref{tab:constructiond}. (These properties, and the operation
frequency $\w = $1 kHz, were used to generate the optimization curves
in section ~\ref{sec:opt}).  To reduce the likelihood of 
acoustic emission, a value for the elastic limit of stainless steel 
equal to 1/3 the tabulated value is assumed.
{\squeezetable
\begin{table}
\caption{Properties of spring wires used in \tvi \ model and construction.}
\label{tab:construction}
\begin{tabular}{lr}\hline
\multicolumn{1}{c}{\it Parameter} & \multicolumn{1}{c}{\it Value}\\ \hline
Material & Type 304 Stainless Steel \\
Density, $\rho_{w}$ & $8.05 \times 10^{3} \mbox{\ kg/m}^{3}$ \\
Elastic Modulus, $E$ & $2.8 \times 10^{11} \mbox{\ N/m}^{2}$ \\
Elastic Limit \tablenote{To limit acoustic emission, the value 
  actually used in the model and design is 1/3 the tabulated value:
  $7 \times 10^{8} \mbox{\ N/m}^{2}$} , $s$ 
  & $2.1 \times 10^{9} \mbox{\ N/m}^{2}$ \\
Total Length (along single row of posts), $L$ & 16.2 cm \\
Length per Stage, $l$ & 2.7 cm \\
Diameter, $d_{w}$ & stage 1:  $45 \ \mu\mbox{m}$ \\
 & stages 2,3: $40 \ \mu\mbox{m}$ \\
 & stages 4,5: $30 \ \mu\mbox{m}$ \\
 & stage 6: $25 \ \mu\mbox{m}$ \\
\end{tabular}
\end{table}
}
\begin{table}
\caption{Properties of mass disks used in \tvi \ model construction.}
\label{tab:constructiond}
\begin{tabular}{lr}\hline
\multicolumn{1}{c}{\it Parameter} & \multicolumn{1}{c}{\it Value}\\ \hline
Material & Brass \\
Density, $\rho_{d}$ & $8.5 \times 10^{3} \mbox{\ kg/m}^{3}$ \\
Radius, $R$ & 1.9 cm \\
Thickness, $t$ & 1.3 cm \\
\end{tabular}
\end{table}

The required degree of vibration isolation can be roughly estimated 
as the ratio of the force exerted by our
gram--mass oscillator resonating at 1 kHz to the gravitational force
between it and a similar oscillator spaced 1 mm away. This is roughly
$10^{12}$, or 240 dB.

Reading off Fig.~\ref{fig:attenuation}, 240 dB requires a stack of at
least five stages.  A six--stage design is chosen to conserve on labor
and maintenance, which with fine wires can be delicate and 
time--consuming.  Fig.~\ref{fig:transverse} indicates that any 
transverse spring modes  in a six--stage stack will 
have frequencies greater 5 kHz, safely above the operating range. Finally,
Fig.~\ref{fig:wire} shows that a six--stage stack requires spring wire
diameters ranging from about 30 to 80 $\mu$m.

The expected loss in attenuation due to the sub--optimal disk
thickness can be compensated by choosing smaller wire gauges
than specified in Fig.~\ref{fig:wire}.  The design is optimized assuming an
elastic limit below the actual maximum, so the final design uses wire
diameters ranging from 25 to 45 $\mu$m.

\subsection{Assembly}
The choice of six stages fixes the length of each wire per stage to
$l = L/n = 16.2/6 = 2.7$ cm.  The disks have dimensions $R = l/
\sqrt{2} = 1.9$ cm, $t = l/2 = 1.3$ cm, and are easily machined
from brass stock. Seven disks are machined, with the first intended to
attach to the support structure.

Holes are drilled and tapped into the center plane of each disk at
60 degree intervals to provide for the six 6-32 screws which serve
as wire attachment posts.  Additional holes are drilled into the top
and bottom surfaces of the first and seventh disk, for mounting to the
support structure and for attaching the isolated instruments. 

\begin{figure}[hbtp]
\begin{center}
\epsfxsize=7.0cm \epsfbox{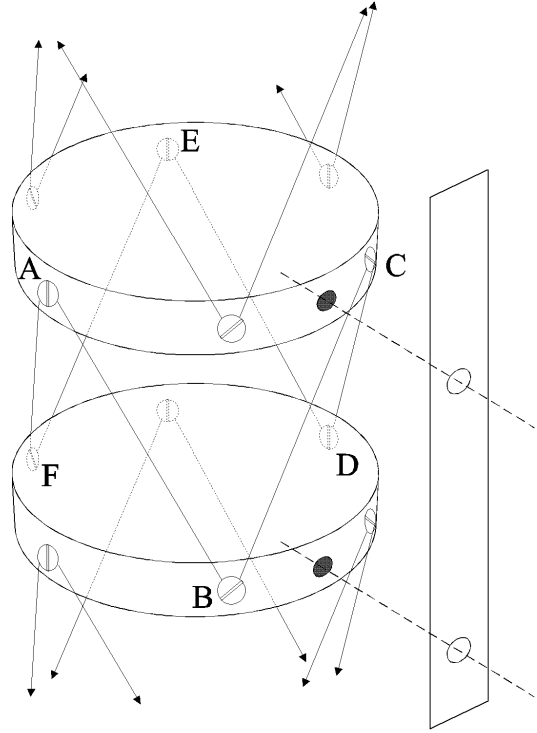}
\caption{Wiring procedure for construction of multi--stage \tvi \ (not
  to scale), showing a segment of one of the three temporary brass 
  holding strips.}
\label{fig:wiring}
\end{center}
\end{figure}

Since the attachment of the wires is a delicate procedure, 
the seven disks must first be mounted securely.  For this 
purpose, additional holes are 
drilled into the center plane of each disk, at 120 degree intervals.  
Three thin brass strips, each with seven holes spaced 1.9 cm apart, 
can then be attached to the stack using the extra holes, firmly
fixing the disks in a vertical stack (Fig.~\ref{fig:wiring}).   

\begin{figure}[hbtp]
\begin{center}
\epsfxsize=7.5cm \epsfbox{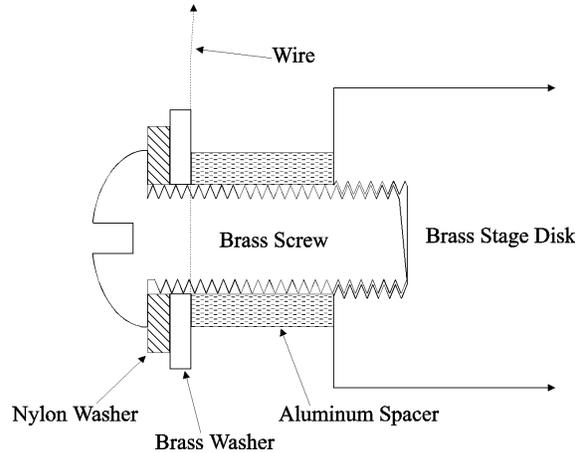}
\caption{Cross--section of wire post construction on edge of stage
  disk (not to scale).}
\label{fig:post}
\end{center}
\end{figure}

With the strips in place, the wire attachment posts are assembled
using the remaining holes in each disk.  Each post consists of a
vented aluminum spacer, a thin brass washer, and a thin nylon washer,
secured to the disk with a vented screw (Fig.~\ref{fig:post} and 
Table~\ref{tab:constructiona}). 

The posts are designed so that a single strand of spring wire can be
used to support each stage, rather than six separate strands of
wire.  For example, in Fig.~\ref{fig:wiring}, one end of a 16.2 cm
piece of wire
is attached between the brass and aluminum washers at point A, and the
other end is attached to a small weight. The wire is
then woven between the washers at points B through F successively, and
re-attached at point A.  During this operation, as soon as the wire is 
woven around a particular post, the weighted end is left to hang while 
the post screw is tightened, thereby insuring uniform wire tension
between each post. The nylon washer on each post is lubricated with
vacuum pump oil, so that it does not rotate the brass washer (and
thereby apply torque to the wire) as the screw is tightened. 

\begin{figure}[hbtp]
\begin{center}
\epsfysize=10.0cm \epsfbox{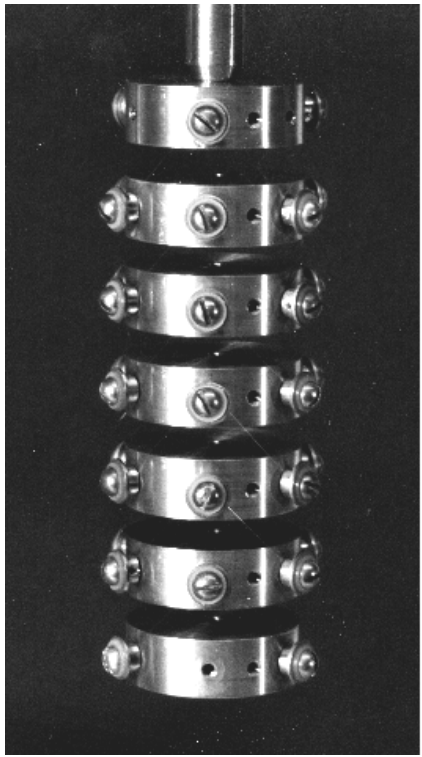}
\caption{Photo of completed \tvi.}
\label{fig:photo}
\end{center}
\end{figure}

As per the optimization procedure, wires of smaller gauges are used to
connect successively lower stages, as specified in
Table~\ref{tab:construction}.  After all post screws are tightened on
each stage, the brass strips are removed and the finished isolator is
ready for use (Fig.~\ref{fig:photo}).
\begin{table}
\caption{Items used in spring wire post construction.}
\label{tab:constructiona}
\begin{tabular}{lcr}\hline
\multicolumn{1}{c}{\it Item} & \multicolumn{1}{c}{\it Material} & 
\multicolumn{1}{c}{\it Dimensions (mm)} \\ \hline
6-32 Screw & Brass & 9.5\\
Spacer & Aluminum & 3.2 $\times$ 3.7 I.D. $\times$ 6.3 O.D. \\
Clamp Washer & Brass & 0.8 $\times$ 3.7 I.D. $\times$ 9.5 O.D. \\
Guide Washer & Nylon & 0.8 $\times$ 3.7 I.D. $\times$ 7.9 O.D. \\
\end{tabular}
\end{table}

\subsection{Single--Stage Test}
The experimental verification of a multi--stage \tvi\ is a challenging
problem due to the difficulty of observing the 
multiple, closely--spaced normal modes and the highly attenuated 
motion of the lower stages.  We limit ourselves here to checking the
natural frequencies of a single stage.

The single--stage isolator is constructed using the materials and 
specifications in 
Tables~\ref{tab:construction}--\ref{tab:constructiona} 
(the diameter of the single wire is 30 $\mu$m). Using the same set 
of design parameters, plus some small refinements to include
the wire posts, a 
finite element analysis is performed with ANSYS. The normal mode 
frequencies are also calculated analytically using formulae similar
to those in Table ~\ref{tab:k}, but with corrections
for the effect of attaching the spring wires at points outside 
the disk radii, and for the additional rotational inertia due to the 
wire posts. These results are summarized in Table~\ref{tab:onestage}. 
\begin{table}
\caption{Natural frequencies for each \dof\ of a single--stage \tvi.}
\label{tab:onestage}
\begin{tabular}{lccc}\hline
\multicolumn{1}{c}{ } & \multicolumn{1}{c}{\it Analytical} 
& \multicolumn{1}{c}{\it ANSYS} 
& \multicolumn{1}{c}{\it Measurement} \\
\multicolumn{1}{c}{\it Motion} & \multicolumn{1}{c}{\it Model (Hz)}
& \multicolumn{1}{c}{\it Result (Hz)} & \multicolumn{1}{c}{\it (Hz)}
\\ \hline
Translation, 1-axis & 56 & 42 & 38\\
Translation, 2-axis & 56 & 42 & 38\\
Translation, 3-axis & 57 & 62 & 65\\
Rotation, 1-axis & 86 & 96 & 98\\
Rotation, 2-axis & 86 & 96 & 98\\
Rotation, 3-axis & 110 & 101 & 102\\
\end{tabular}
\end{table}

As in the multi--stage case in Sec.~\ref{sec:ansys}, the methods are
in better agreement (within 9\%) for translations along and 
rotations about the 
3-axis.  The 10--30\% discrepancies between the other modes arise
because the pure 1- and 2- translations and rotations assumed by 
the analytical model
are only approximations of the actual motion for these modes.  The
ANSYS graphics indicate that these modes involve pendulum--type
motions. The modes labeled ``translations'' in the transverse
directions in Table~\ref{tab:onestage} actually mix the translations
with slight rotations about the transverse axes.  Similarly, ``rotations'' mix
with slight translations, so that the labels in
Table~\ref{tab:onestage} are to some extent misnomers for the case of
the numerical analysis (and for the measurements).

The resonant frequencies of the single--stage \tvi\ are measured with
a PZT transducer.  The transducer is bolted to the stage using one
of the spare holes for the brass assembly strips, and the output is
fed to a spectrum analyzer.  Six frequencies are observed. 
In order to identify an observed frequency with a particular mode, the
spectrum analyzer response is monitored as the sensitive axis of
the PZT is oriented in each direction associated with the predominant
motion of the predicted modes.  The largest signal on the analyzer for
a particular orientation is then recorded as the frequency of the
associated mode.  

\begin{figure}[hbt]
\begin{center}
\epsfxsize=8.5cm \epsfbox{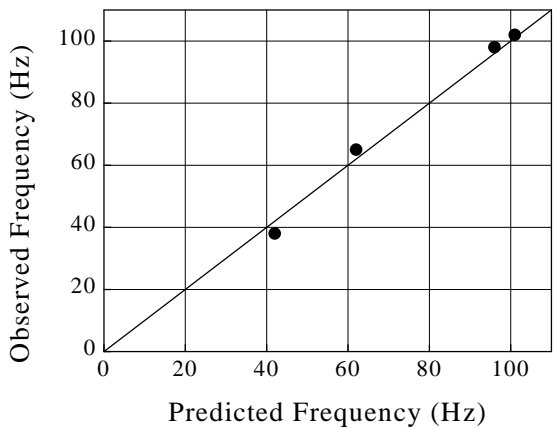}
\vspace{.5cm}
\caption{Measured resonant frequencies for single--stage isolator
  vs. numerical analysis predictions.  The frequencies near 40 and 95
  Hz are two--fold degenerate. Line indicates unit slope.}
\label{fig:fvf}
\end{center}
\end{figure}

Using this procedure, the six frequencies can be identified with 
expected modes.  The measurements are plotted in Fig.~\ref{fig:fvf}, 
against the predictions from the finite element analysis.  
The results agree to within 10\% for
each frequency.

A conservative assumption is that the attenuation of the six--stage
\tvi\ at the operating frequency $\w = 1$ kHz is limited by the 
highest natural frequency. From Fig.~\ref{fig:fvf}, the highest
frequency is $\w_{0} \approx$ 100 Hz.  The estimate of the attenuation is then
$-20 \log{(\w_{0}/\w)^{2n}}$ = 260 dB, in accordance with the model in
Sec.~\ref{sec:design}.
     
\section{Conclusions}
An approximate analytical model for a multi--stage \tvi \ can be
constructed from a one--dimensional spring--mass chain, in which the
spring constants are derived from the geometry of a single stage.  The
full solution to this model makes exact predictions for the normal
modes involving translations along and rotations about the
longitudinal axis of the isolator, as computed with a finite element
analysis.  The model gives good approximations for the transverse
modes as well. 

The high--frequency behavior of an isolator with 
identical (or nearly identical) stages is well characterized by 
the natural frequencies of a
single stage in all \dof, which are easily calculated with the model.
These frequencies depend on the design parameters of the \tvi \ in a
straightforward way and are very useful for the design of a
multi--stage isolator.  

We have used the single--stage frequency formulae to design a
fixed--length \tvi \ for operation at 1 kHz in all \dof.
The model illustrates the importance of selecting stage masses of high
density, and spring materials of low elastic modulus and high elastic
limit. The last quality is important for both the maximization of the
attenuation and for ensuring that the transverse modes of the
spring wires are well above the operating frequency.

The model makes accurate predicitons for the single--frequencies in the
final design, which are in the range of 100 Hz.  With six stages, the
attenuation of our centimeter--scale \tvi \ is estimated at 260 dB.

\end{document}